\documentclass[%
 reprint,
 superscriptaddress,
 amsmath,amssymb,
 aps,
]{revtex4-2}

\usepackage{graphicx}
\usepackage{dcolumn}
\usepackage{bm}
\usepackage[colorlinks=true,citecolor=blue]{hyperref}
\usepackage{color}

\newcommand{\fermion} {\hat c}
\begin{document}

\preprint{APS/123-QED}

\title{Many-body Majorana-like zero modes without gauge symmetry breaking}

\author{V. Vadimov}
\affiliation{QCD Labs, QTF Centre of Excellence, Department of Applied Physics, Aalto University, P.O. Box 13500,FI-00076 Aalto, Espoo, Finland}
\affiliation{MSP Group, QTF Centre of Excellence, Department of Applied Physics, Aalto University, P.O. Box 11000, FI-00076 Aalto, Espoo, Finland}
\affiliation{Institute for Physics of Microstructures, Russian
Academy of Sciences, 603950 Nizhny Novgorod, GSP-105, Russia}

\author{T. Hyart}
\affiliation{Department of Applied Physics, Aalto University, FI-00076 Aalto, Espoo, Finland}
\affiliation{International Research Centre MagTop, Institute of Physics, Polish Academy of Sciences, Al. Lotników 32/46, 02-668 Warsaw, Poland}

\author{J. L. Lado}
\affiliation{Department of Applied Physics, Aalto University, FI-00076 Aalto, Espoo, Finland}

\author{M. M\"ott\"onen}
\affiliation{QCD Labs, QTF Centre of Excellence, Department of Applied Physics, Aalto University, P.O. Box 13500,FI-00076 Aalto, Espoo, Finland}
\affiliation{VTT Technical Research Centre of Finland Ltd., QTF Center of Excellence, P.O. Box 1000, FI-02044 VTT, Finland}

\author{T. Ala-Nissila}
\affiliation{MSP Group, QTF Centre of Excellence, Department of Applied Physics, Aalto University, P.O. Box 11000, FI-00076 Aalto, Espoo, Finland}

\affiliation{Interdisciplinary Centre for Mathematical Modelling, Department of Mathematical Sciences, Loughborough University, Loughborough LE11 3TU, UK}

\date{\today}

\begin{abstract}
Topological superconductors represent one of the key hosts of Majorana-based topological
quantum computing. Typical scenarios for one-dimensional topological superconductivity
assume a broken gauge symmetry associated to a superconducting state.
However,
no interacting one-dimensional many-body system is known to spontaneously break
gauge symmetries.
Here, we show that
zero modes emerge in a many-body system without gauge symmetry breaking
and in the absence of superconducting order. 
In particular, we demonstrate that Majorana zero modes
of the symmetry-broken superconducting state are continuously connected to these
zero-mode excitations, demonstrating that zero-bias anomalies may emerge in the absence
of gauge symmetry breaking. We demonstrate that these many-body
zero modes share the robustness features of the Majorana zero modes
of symmetry-broken topological superconductors.
We introduce a bosonization formalism to analyze these excitations
and show that a ground state analogous to a topological
superconducting state can be analytically found in a certain limit.
Our results demonstrate that robust Majorana-like zero modes
may appear in a many-body systems without gauge symmetry breaking, thus introducing
a family of 
protected excitations with no single-particle analogs.
\end{abstract}

\maketitle

\section{Introduction}
Superconductivity in topological quantum materials has become one of the most fertile topics in modern condensed matter physics~\cite{qi2011,sato2017}.
The search for topological superconductors has been motivated by the emergence of topological excitations,
known as Majorana zero modes~\cite{RevModPhys.87.137}, and by their potential for topological quantum
computing~\cite{Ivanov, kitaev2003fault,RevModPhys.80.1083, Sato09, PhysRevLett.104.040502, PhysRevB.82.214509,   Hyart13, Karzig17, Lutchyn2018,  Beenakker20}.
A variety of solid-state materials have been explored in recent years with the
goal of engineering Majorana bound states, including superconducting nanowires~\cite{PhysRevLett.105.077001,oreg2010, PhysRevB.84.144522, Mou12, PhysRevLett.110.126406, PhysRevB.87.241401, PhysRevLett.110.126406,  PhysRevB.87.024515, PhysRevB.87.241401,  PhysRevX.8.031041,PhysRevX.3.041017}, atomically engineered
chains~\cite{Choy11,PhysRevB.88.020407,Nad14,PhysRevB.90.060507}, topological insulators~\cite{PhysRevLett.100.096407, fu2009josephson, PhysRevLett.109.056803, yacobyNP2014, PribiagNatNano2015}, phase-controlled Josephson junctions~\cite{HechenNature2019, FornieriNature2019}, helical quantum Hall edge states of graphene~\cite{PhysRevX.5.041042} with controllable magnetic~\cite{Yang20, Veyrat20} and superconducting gaps, antiferromagnetic topological superconductors~\cite{PhysRevLett.114.056403,PhysRevLett.121.037002}, and
van der Waals heterostructures~\cite{2020arXiv200202141K}.
These different platforms rely on
the engineering of a specific kind of an effective $p$-wave superconducting state,
the non-trivial topological properties of which give rise to the emergence of Majorana
zero modes~\cite{kitaev2001}.

\begin{figure}[t!]
    \centering
    \includegraphics[width=0.8\linewidth]{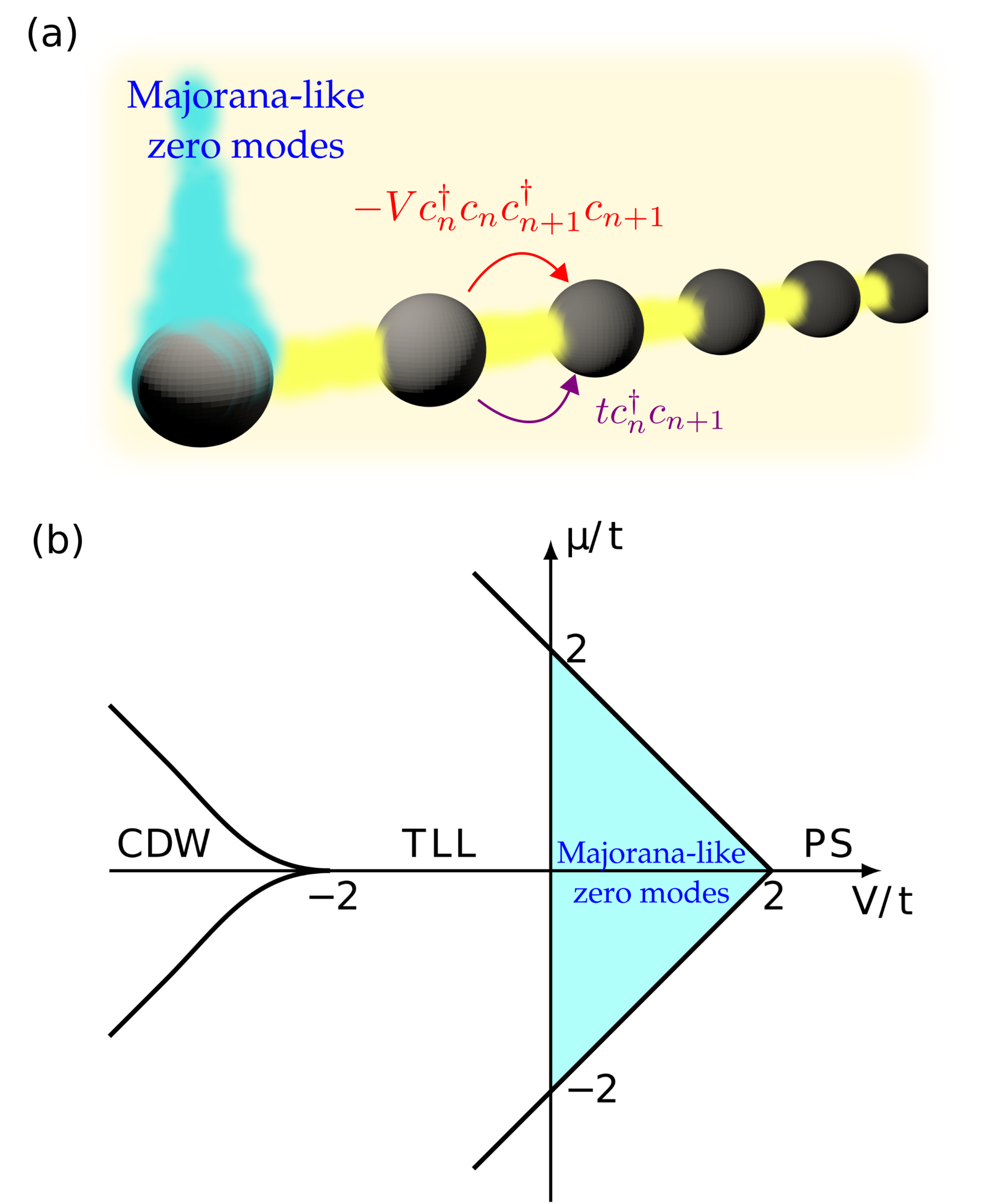} 
    \caption{(a) Schematic illustration of the interacting one-dimensional model we study. The
    Majorana-like zero energy mode at the left edge of the chain is highlighted with cyan color. The fermionic creation and annihilation operators at site $n$ (grey sphere) are denoted by $\fermion_n^\dagger$ and $\fermion_n$, respectively. The strength of the particle hopping between neighboring sites (yellow color) is given by $t$, and $V$ describes the strength of the  interactions between the fermions.
    (b) The phase diagram of the Hamiltonian~\eqref{eq:fermi-hamiltonian} at zero temperature. Here, $\mu$ denotes the chemical potential. Regions corresponding to a phase separation (PS), Tomonaga--Luttinger liquid (TLL), and charge density wave (CDW) appear in addition to Majorana-like zero modes (cyan color). 
    }
    \label{fig:diagram}
\end{figure}

Majorana bound states and unconventional superconductors in general rely on a single-particle description of the effective excitations. In particular, conventional proposals
for Majorana bound states in topological superconductors
rely on explicitly broken gauge symmetry which is associated with the existence of a non-zero
superfluid order parameter~\cite{kitaev2001}.

In this scenario, the existence of Majorana bound states
in the presence of particle--particle interactions has been
established in terms of a renormalization of the
single-particle mean-field
parameters~\cite{Fisher11,PhysRevB.85.140508,PhysRevB.88.161103,PhysRevLett.107.036801,PhysRevB.89.220504}. 
Although additional subtle many-body effects are prone to appear in this regime~\cite{PhysRevX.4.031051,PhysRevLett.109.146403,PhysRevB.84.014503,PhysRevB.101.125431}, the Majorana zero
remain to exist.

However, the scenario for systems lacking symmetry breaking is distinctively different~\cite{PhysRevB.91.235309,PhysRevB.84.195436,PhysRevB.84.144509,PhysRevB.97.165114,PhysRevB.92.041118,PhysRevX.7.041048,PhysRevLett.120.057001,PhysRevB.98.214501,PhysRevLett.114.100401,PhysRevLett.113.267002}.
Namely, particle--particle interactions cannot be reinterpreted as a renormalization of single-particle
terms as in a symmetry broken state.
For typical single-particle models of topological superconductivity,
the pairing term explicitly breaks the gauge symmetry since the term is not
$U(1)$ gauge symmetric. 
A finite pairing term is a hallmark
of superconductivity, and thus such symmetry breaking
is natural for proposals that involve 
three-dimensional superconductors~\cite{Alicea2012}.
However, for a purely one-dimensional (1D) system, the situation
is dramatically different, as spontaneous symmetry breaking
with finite pairing does not take place~\cite{PhysRev.158.383,PhysRevLett.17.1133}.
In particular, interacting 1D models have a ground state
that is gauge symmetric with a vanishing expectation value of superconducting
pairing~\cite{PhysRev.158.383,PhysRevLett.17.1133},
and consequently their effective single-particle Hamiltonian does not host
Majorana zero modes~\cite{kitaev2001,Alicea2012}.
Thus, whether or not
Majorana zero modes may appear in the
absence of gauge symmetry breaking is a major outstanding question.

In this work, we demonstrate that robust zero modes
appear in a 1D many-body model without gauge symmetry breaking.
The model we focus on would give rise to a topological superconductor
at the mean-field level if the gauge symmetry were explicitly broken.
We demonstrate that no such gauge symmetry
breaking is required for the emergence
of Majorana-like zero modes, establishing a peculiar
paradigm of topological quantum many-body excitations
with no single-particle analog.
Despite their fundamental
differences to Majorana zero modes, we demonstrate
that these two types of many-body excitations share many properties,
including robustness to perturbations and disorder.

Our manuscript is organized as follows.
In Sec.~\ref{sec:model}, we introduce the many-body model, highlighting the emergence
of zero-mode resonances. In Sec.~\ref{sec:pert}, we demonstrate the robustness
of the zero modes to a variety of perturbations. 
In Sec.~\ref{sec:majo}, we show the connection between these resonant zero modes
and Majorana bound states.
In Sec.~\ref{sec:bos}, we demonstrate the 
emergence of the edge modes from a continuum bosonization formalism. 
Section~\ref{sec:con} summarizes our results and Appendices~\ref{sec:critical}--\ref{sec:persistentcurrent} discuss some technical details such as critical points, Green's functions, and persistent current in a ring.

\section{Zero modes in quantum many-body chains}
\label{sec:model}

We study a 1D chain of $L$ spinless fermions with interactions between the neighboring sites
as illustrated in Fig.~\ref{fig:diagram}(a). The system is described by the following Hamiltonian:
\begin{multline}
    \hat H_I = -t \sum\limits_{j=1}^{L-1} \left(\fermion_{j+1}^\dag \fermion_j + \fermion_j^\dag \fermion_{j+1}\right)   - \mu \sum\limits_{j=1}^L \fermion_j^\dag \fermion_j \\ - V \sum\limits_{j=1}^{L-1} \left(\fermion_{j+1}^\dag \fermion_{j+1} - \frac{1}{2}\right)\left(\fermion_j^\dag \fermion_j- \frac{1}{2}\right),
    \label{eq:fermi-hamiltonian}
\end{multline}
where $t$ is the strength of the particle hopping between neighboring sites, $\fermion_j^\dagger$ and $\fermion_j$ are the fermionic creation and annihilation operators at site $j$, respectively, $\mu$ is the chemical potential, and $V$ is the strength of the interactions between the fermions. 
Such a system can be mapped onto a spin-1/2 anisotropic XXZ chain in a longitudinal field using the Jordan--Wigner transformation, resulting in
$$
    \hat H_I = \sum\limits_{j=1}^{L-1}\left[2t \left(\hat s^x_j \hat s^x_{j+1} + \hat s^y_j \hat s^y_{j+1}\right) - V \hat s^z_{j} \hat s^z_{j+1}\right] - \mu \sum\limits_{j=1}^L \hat s^z_j,
    \label{eq:spin-hamiltonian}
$$
where $\{\hat{s}_j^\alpha\}_{\alpha=x, y, z}$ denote the spin-$1/2$ operators for site $j$.
This model is integrable by the means of the Bethe ansatz~\cite{alcaraz1987,sklyanin1988}. The resulting phase diagram at zero temperature and otherwise in the thermodynamic limit is shown in Fig.~\ref{fig:diagram}(b)~\cite{mikeska2004one}. We focus on the region of the diagram corresponding to the attractive interactions between the fermions $V > 0$, where two different phases exist. Phase separation takes place at $|\mu| > 2 t - V$ where the ground sate corresponds, depending on the sign of $\mu$, to the vacuum state or to the completely filled band. 
In this phase, zero-energy modes which mix the number of particles are known to exist~\cite{fendley2016strong}. The other phase corresponds to the Tomonaga--Luttinger liquid~\cite{tomonaga1950,luttinger1963,haldane1981} which is our main focus here. Remarkably, we have been able to explicitly construct the ground state at the critical point $V = 2t$, $\mu=0$ as shown in Appendix~\ref{sec:critical}. At this point it appears to be $L+1$ times degenerate, where the different degenerate states correspond to the different numbers of particles. 

Treated within the mean-field approximation such a model gives rise to the well-known Kitaev model described by the Hamiltonian
\begin{multline}
    \hat H_K = -t \sum\limits_{j=1}^{L-1} \left(\fermion_{j+1}^\dag \fermion_j + \fermion_j^\dag \fermion_{j+1}\right)   - \mu \sum\limits_{j=1}^L \fermion_j^\dag \fermion_j + \\ \sum\limits_{j=1}^{L-1} \left(\Delta_{j} \fermion_{j+1}^\dag \fermion_j^\dag + \Delta_j^\ast \fermion_j \fermion_{j+1}\right),
\end{multline}
where the superconducting order parameter 
$\Delta_j$ is determined self-consistently as $\Delta_j = V \langle \fermion_j \fermion_{j+1}\rangle$. The Kitaev model spontaneously breaks the gauge symmetry which is present in the original model~\eqref{eq:fermi-hamiltonian}. For $L \gg t / |\Delta|$ this model hosts Majorana zero modes localized at the ends of the chain~\cite{kitaev2001} whereas the other excited states are separated from the ground state by $|\Delta|$ in the bulk of the chain. 

\subsection{Local density of states}
\begin{figure}[t!]
    \centering
    \includegraphics[width=\linewidth]{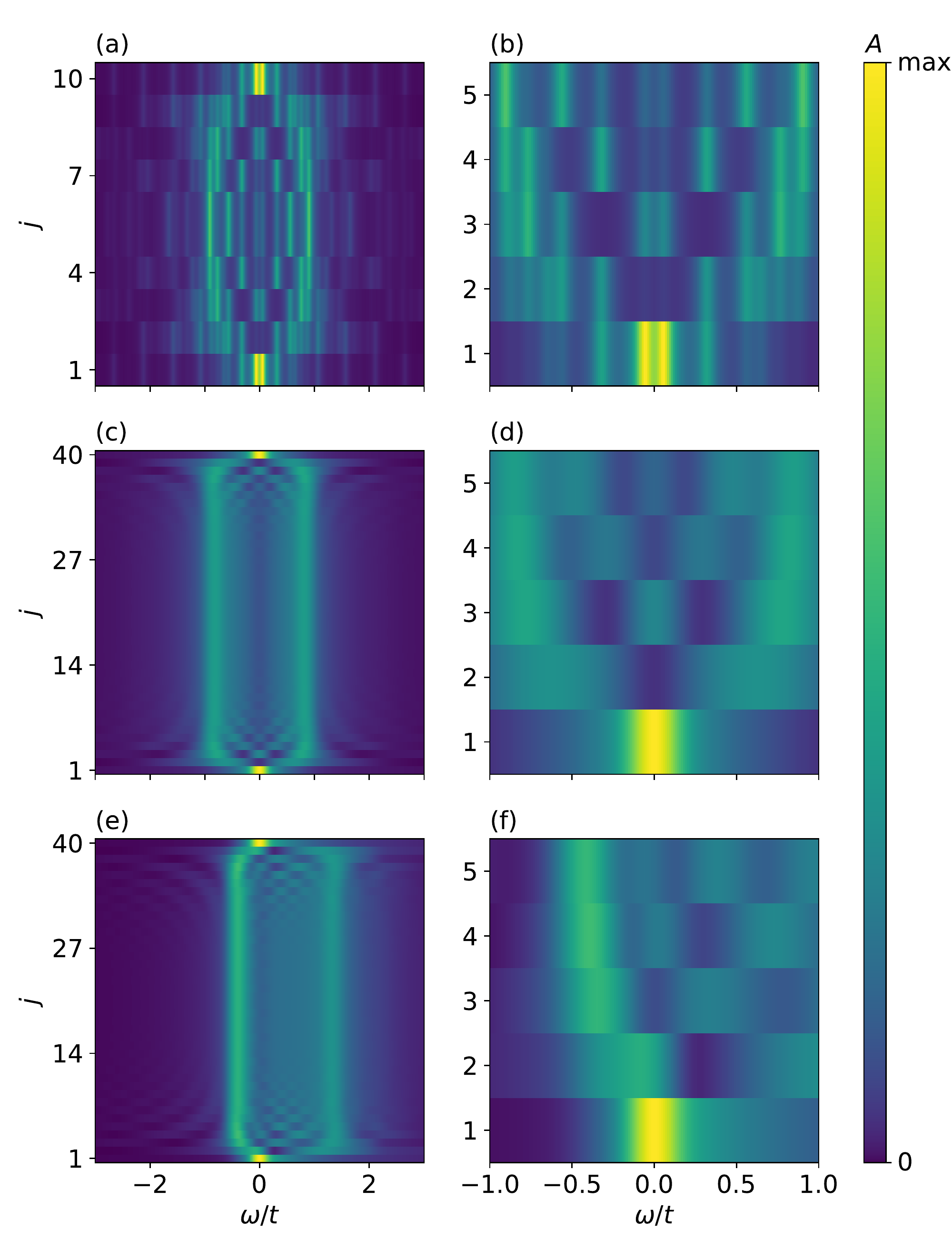} 
    \caption{Local density of states as a function of frequency and the site of the chain for (a,b) $L = 10$, $\mu = 0$ (c,d) $L = 40$, $\mu = 0$ (e,f) $L = 40$, $\mu = -0.2 t$. Panels (b,d,f) show the maps of the local density of states $A(j, \omega)$ zoomed to the edge of the chain from the corresponding panels (a,c,e) . The interaction strength is $V = 1.5t$. The panels (a,b) have been obtained using exact diagonalization of the Hamiltonian~\eqref{eq:fermi-hamiltonian} and the others are obtained using the KPM-MPS method. 
    }
    \label{fig:ldos}
\end{figure}
\begin{figure}
    \centering
    \includegraphics[width=\linewidth]{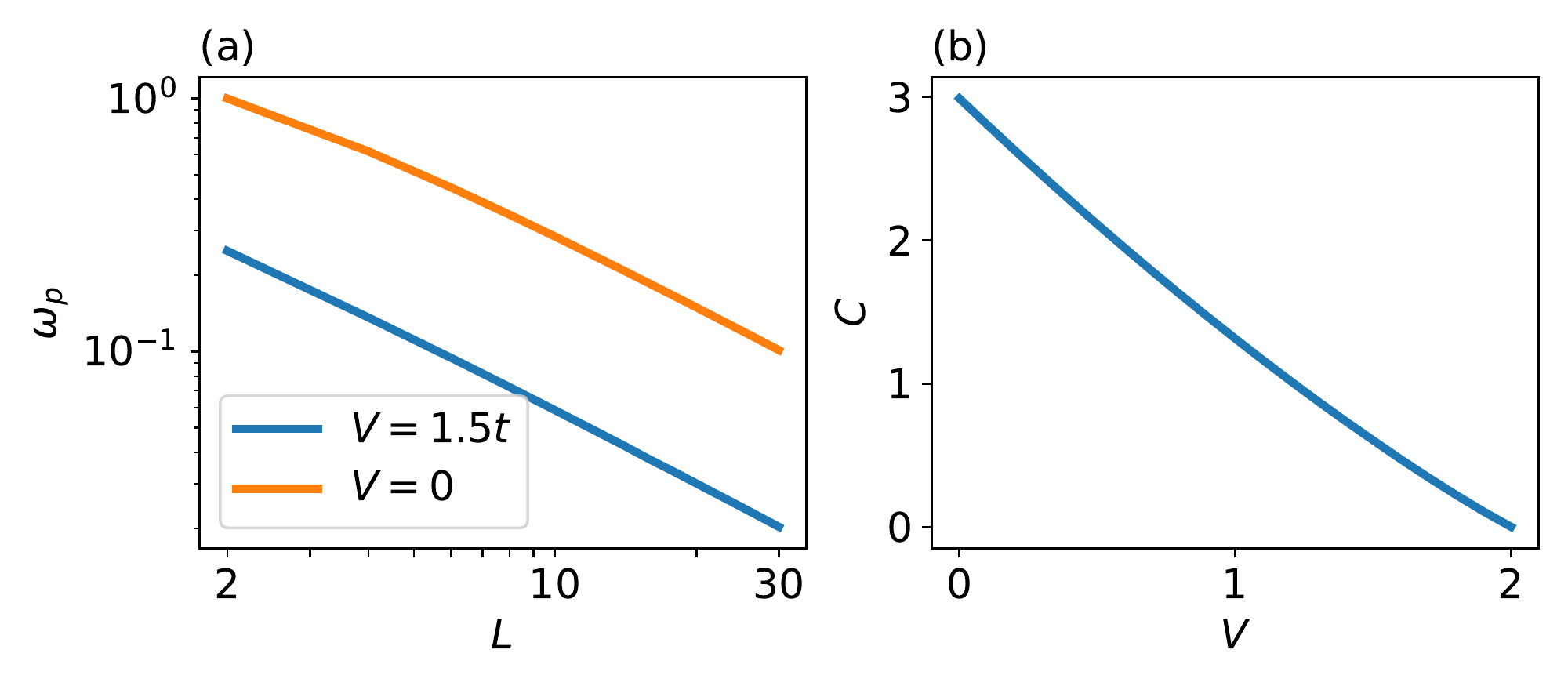}
    \caption{(a) Scaling of the peak splitting with $L$. (b) Dependence of the prefactor $C$ on the interaction strength $V$. The calculations are performed using MPS algorithm with $\mu = 0$.}
    \label{fig:scaling}
\end{figure}
The local density of states, or the spectral function, of the chain is defined as
\begin{multline}
    A(j,\omega) = \langle \Psi_0 |\fermion_j \delta(\omega -\hat H + E_0)\fermion_j^\dag +\\ \fermion_j^\dag\delta(\omega + \hat H - E_0)\fermion_j |\Psi_0\rangle = \\
    \sum\limits_m \left[|\langle \Psi_0| \fermion_j |\Psi_m\rangle|^2 \delta(\omega - E_m + E_0) +\phantom{a_j^\dag}\right. \\ \left. |\langle \Psi_0| \fermion_j^\dag |\Psi_m\rangle|^2 \delta(\omega - E_0 + E_m)\right],
    \label{eq:ldos}
\end{multline}
where $|\Psi_0\rangle$ is the ground state with energy $E_0$ and $|\Psi_m \rangle$ are all the eigenstates of the system corresponding to energies $E_m$. The spectral function 
can be evaluated using exact diagonalization of the Hamiltonian for the short chains or using a kernel polynomial method with matrix product states
(KPM-MPS)~\cite{RevModPhys.78.275,PhysRevB.91.115144,PhysRevResearch.1.033009,PhysRevResearch.2.023347,2020arXiv200714822F,ITensor,dmrgpy} for the reasonably long chains. 

The results of our calculations are shown in Fig.~\ref{fig:ldos}. We observe
clear zero-bias peaks at the edges of the chain, corresponding to Majorana-like zero modes. We find that these zero-energy
peaks appear
for all considered chemical potentials $\mu$. 
Similarly to Majorana edge modes,
the peaks split for short chains as shown in Fig.~\ref{fig:ldos}(b), stemming from the hybridization of the excitations at the opposite edges.
With increasing chain length $L$, the edge peaks move towards zero energy, constituting a zero-energy resonance in the
limit $L\to\infty$. Next, we systematically examine how the splitting of these edge modes depends
on the system size. Interestingly, their 
behavior is different from 
that of Majorana zero modes in topological superconductors.

\subsection{Peak scaling}

The nature of the above-found edge modes can be studied by inspecting the scaling properties of the peak splitting. 
From the series expression in Eq.~\eqref{eq:ldos}, we observe that the the peaks peaks are located at energies
$\pm [E_0(N_0\pm 1) - E_0]$, where $E_0(N)$ is the energy of the ground state in the subspace with $N$ particles and $N_0$ is the number of particles in the global ground state of the system. Thus  the peak splitting $\omega_{\rm p}$ appears to be equal to $[E_0(N_0 + 1) + E_0(N_0 - 1)] / 2 - E_0$ and its dependence of the chain length $L$ is shown in Fig.~\ref{fig:scaling}(a). For comparison, Fig.~\ref{fig:scaling}(a) also shows the same parameter for a non-interacting chain, which corresponds to the level spacing at the Fermi energy. One can see that the splitting scales as $C(V) / L$, which is in contrast to the mean-field case where the splitting of the Majorana peaks decays exponentially with the length of the chain. 
In the case of conventional Majorana states, the exponential dependence arises because of two reasons. First, the bulk of the
system has a gap stemming from the finite pairing. As Majorana zero modes are 
located inside the gap in the bulk of
the system, they need to decay exponentially, which gives rise to hybridization
between zero modes that decays exponentially with the system size. Secondly, the induced superconductivity may be thought as arising from coupling to an infinite superconductor (which does not have charging energy). Since the Majorana wire can exchange Cooper pairs with the infinite superconductor, there is no energy cost for adding particles to the system. As a result of these two effects, the states with $N_0$ and $N_0 \pm 1$ particles are degenerate up to the hybridization energy, which decays exponentially with the
system size. In stark contrast, the present system is different from the conventional Majorana states in both ways. First, it does not have a gap
stemming from the pairing, and the bulk remains gapless. As a result, the Majorana-like
resonances do not have an exponential localization in the edge but rather power-law,
and thus, the hybridization energy decays as a power-law with the system size. Secondly, we are considering a finite system which cannot exchange particles with an infinite superconductor, so that adding particles to the system costs energy $\propto 1/L$, and therefore a splitting $\omega_p \propto 1/L$ would be obtained independently on the type of localization of the end modes (see also Sec.~\ref{sec:bos}). Both contributions are fully included in our calculations.
Dependence of the scaling coefficient $C(V)$ on the interaction strength is shown
in Fig.~\ref{fig:scaling}(b). One can see that it decays almost linearly to $0$ at the critical
point $V = 2t$.

\section{Robustness of the zero modes to perturbations}
\label{sec:pert}

\begin{figure}[t!]
    \centering
    \includegraphics[width=\linewidth]{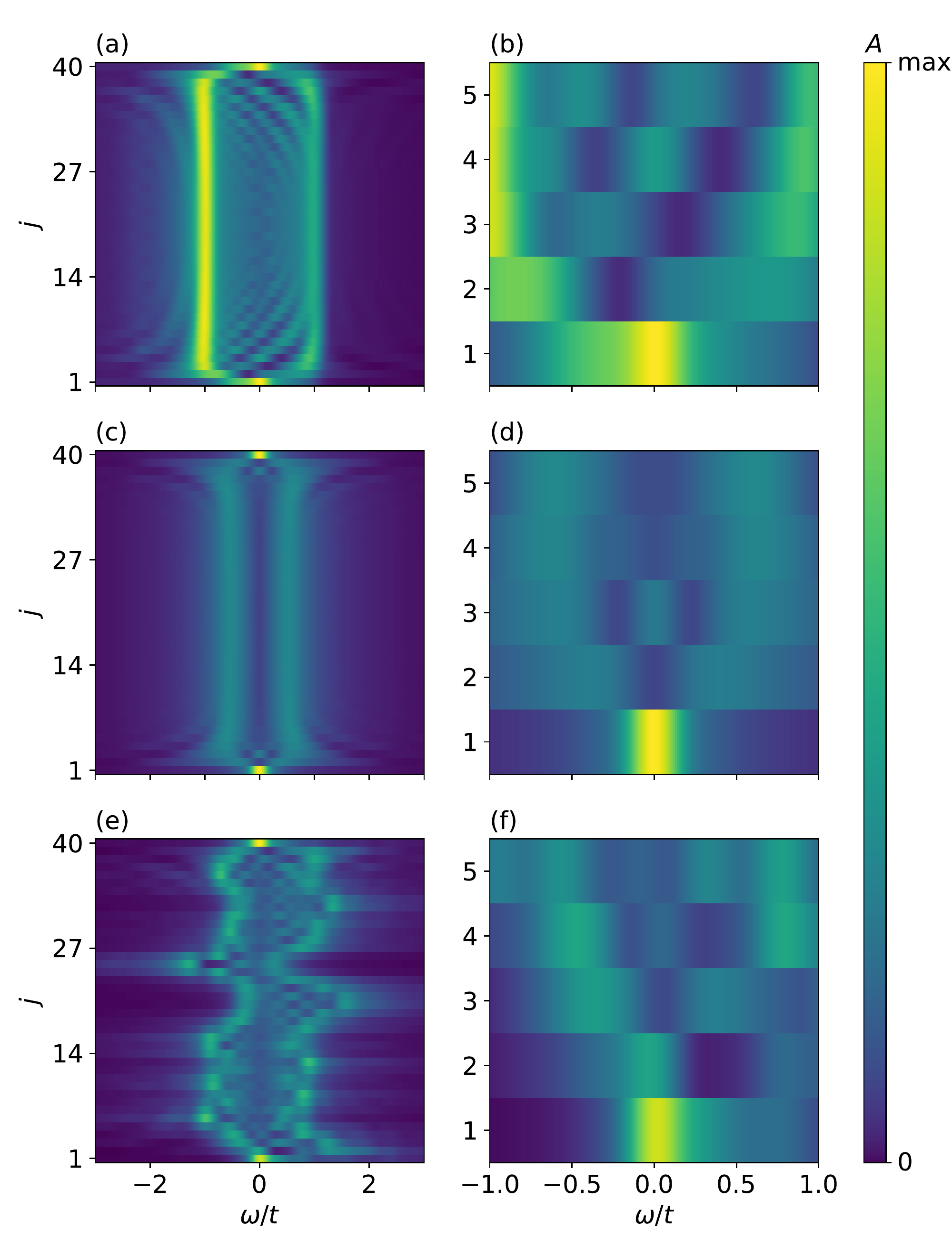} 
    \caption{The local density of states as a function of frequency and the site of the chain for perturbed systems. The perturbations are (a,b) second-neighbor hopping, (c,d) second-neighbor interaction, and (e,f) on-site disorder. 
The parameters are $L = 40$, $V = 1.5t$, $\mu = 0$, (a,b) $t' = -0.2 t$, (c,d) $V'=0.4 t$, (e, f) $W = 0.5 t$.
It can be seen that in all the cases there are strong zero-modes at the edge demonstrating the robustness of these states.
    }
    \label{fig:perturb}
\end{figure}

A paradigmatic property of topological states in general, and Majorana bound states in particular,
is their robustness towards perturbations in the Hamiltonian.
To this end we study robustness of the peaks with respect to perturbations that break the integrability of the model. We consider three different types of perturbations: second-neighbor hopping, second-neighbor interactions, and on-site disorder. These perturbations are described by the following Hamiltonians, respectively:
\begin{gather}
    \hat H_\mathrm{2nh} = -t'\sum\limits_{j=1}^{L-2} \left(\fermion_{j+2}^\dag \fermion_j + \fermion_j^\dag \fermion_{j+2}\right); \\
    \hat H _\mathrm{2ni} = -V' \sum\limits_{j=1}^{L-2} \left(\fermion_{j+2}^\dag \fermion_{j+2} - \frac{1}{2}\right)\left(\fermion_j^\dag \fermion_j - \frac{1}{2}\right), \\
    \hat H_{d} = \sum\limits_{j=1}^L u_j \fermion_j^\dag \fermion_j,
\end{gather}
where $t'$, $V'$ are the parameters controlling the second neighbor hopping and interactions, respectively, and $u_j$ is the on-site potential which is a random number with amplitude $W$ in the range $[-W, W]$. The distributions of local density of states for these types of perturbations are shown
in the Fig.~\ref{fig:perturb}. For all cases here we find that weak perturbations do not affect the existence of the edge states.

\section{Connection to a topological symmetry broken superconductor}
\label{sec:majo}

\begin{figure}[t!]
    \centering
    \includegraphics[width = \linewidth]{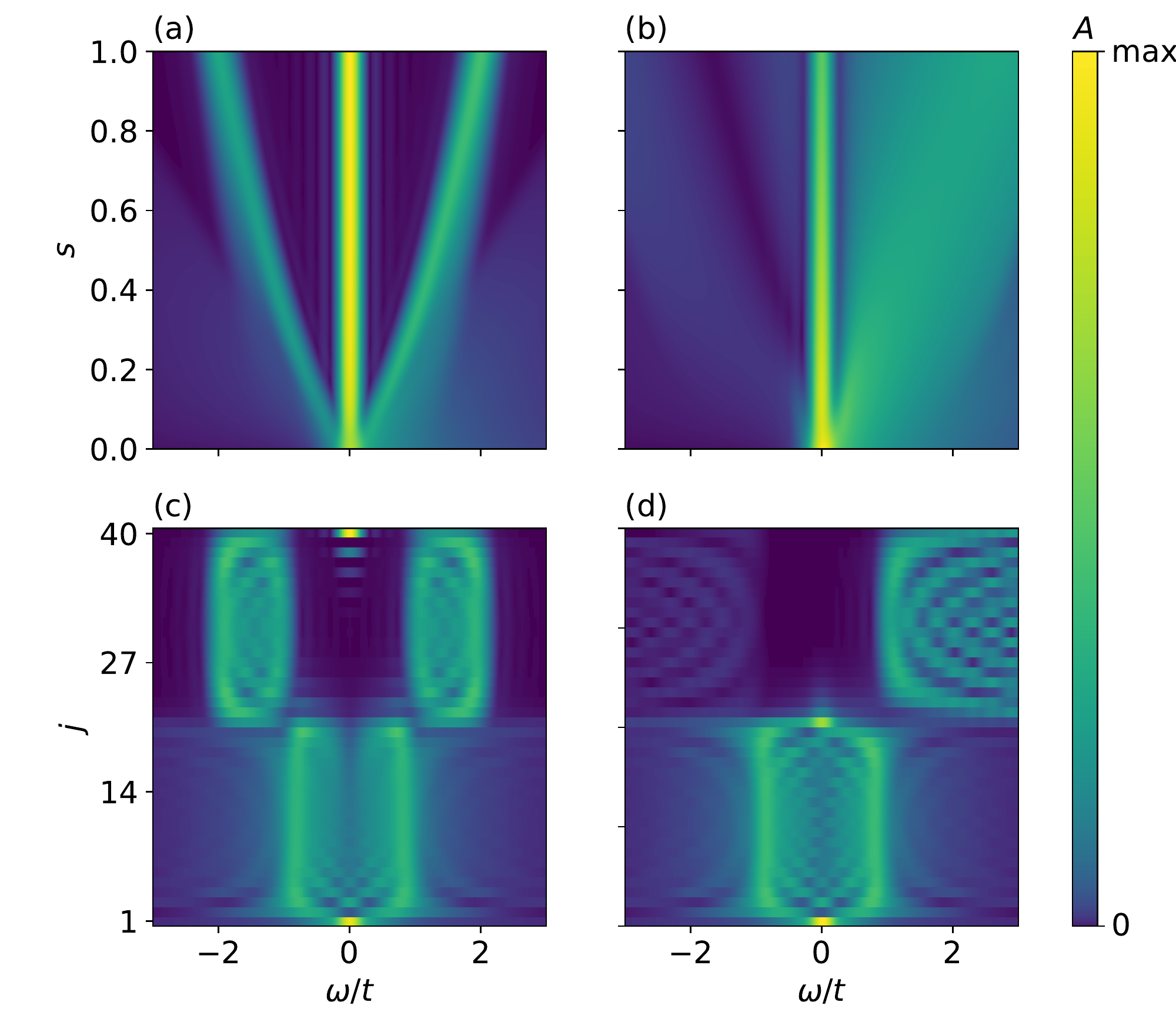} 
    \caption{(a,b) Local density of states at the first site as a function of parameter $s$ and frequency calculated for the Hamiltonian~\eqref{eq:parametric}. Here $L = 40$, $V = 1.5 t$, $\Delta = t$, (a) $\mu =-0.2t$, $\tilde \mu(s) = 0$ and (b) $\tilde\mu(s)$ is adjusted in such a way that $\langle N \rangle(s)/L = 1/4$.
    Panels (c,d) show the local density of states for each site in a heterosjunction
    between a topological superconductor and the interacting model (c),
    and between a trivial superconductor and the interacting model (d). The parameters are $L' = 20$, $\Delta = 0.5 t$, $\mu = 0$, (c) $\mu' = 0$, (d) $\mu' = 3$.}
    \label{fig:connection}
\end{figure}

So far we have focused on the quantum disordered state, showing that the interacting gapless gas shows topological
excitations sharing the robust properties of Majorana zero modes. While such similarity is suggestive, a stronger
insight in the relation between the two can be obtained by demonstrating that the two types of excitations
can be smoothly connected.
In this section we demonstrate that the Majorana-like zero modes can be transformed into
true Majorana bound states by performing an adiabatic connection between the two.
For this purpose we introduce a parametric family of Hamiltonians
\begin{equation}
    \hat H(s) = (1-s) \hat H_I + s \hat H_K - \tilde \mu(s) \hat N,~0 \leqslant s \leqslant 1,
    \label{eq:parametric}
\end{equation}
which smoothly transform from the interacting chain to a topological superconductor
with gauge symmetry breaking by changing of the parameter $s$. 
In particular, for $s=0$ Eq. (\ref{eq:parametric}) becomes the many-body Hamiltonian studied in previous sections,
whereas for $s=1$ Eq. (\ref{eq:parametric}) is a single-particle Hamiltonian
for a topological superconductor. Therefore, changing the parameter $s$ allows tracking the evolution from the Majorana-like
modes without symmetry breaking to Majorana modes with symmetry breaking.
The chemical potential $\tilde \mu(s)$ is defined to control
the total number of particles throughout the path. 

The spectral function at the edge of the chain as a function of the parameter $s$ is shown
in Fig.~\ref{fig:connection}. When the chemical potential $\mu$ does not exceed the transition
value $2t - V$ in absolute value and does not depend on the parameter $\tilde \mu(s) = 0$, the bulk
gap exactly closes at $s = 0$, i.e. in the fully interacting model (see Fig.~\ref{fig:connection}(a)). 
Another way to connect the fully interacting model to the Kitaev chain is to keep the mean number of
particles $\langle N \rangle(s)$ constant with varying the parameter $s$. This case is
shown in Fig.~\ref{fig:connection}(b), which is qualitatively similar to the case of the constant chemical potential.

We would like to emphasize that the connection between the Majorana-like modes here
and the symmetry broken Majorana zero modes
in topological superconductors has some important consequences. In particular,
the Majorana-like modes
here will also appear as a zero-bias anomaly in non-superconducting states, similar to Majorana
zero modes. In this way, both modes would have similar signatures when probed with scanning tunnel microscopy,
appearing as a zero-bias anomaly localized at the edge. These states will however coexist
with a gapless background of edge excitations in the bulk.
Finally, it is worth mentioning that due to the gapless nature of the bulk excitations,
information decoherence in these Majorana-like modes can be different than in conventional
Majorana bound states~\cite{PhysRevB.85.174533,PhysRevB.86.085414,PhysRevB.85.121405}. 

The equivalence between the zero modes of the interacting model and conventional Majorana
zero modes can be further emphasized by studying a heterojunction between the interacting 
model and a topological (trivial) superconductor. The model Hamiltonian of such heterojunction has the following form:
\begin{multline}
    \hat H = -t\sum\limits_{j=1}^{L-1}\left(\fermion_{j+1}^\dag \fermion_j + \fermion_{j}^\dag \fermion_{j+1}\right) -\\ V \sum\limits_{j=1}^{L'-1} \left(\fermion_{j+1}^\dag \fermion_{j+1}-\frac{1}{2}\right) \left(\fermion_{j}^\dag \fermion_j - \frac{1}{2}\right) +\\ \Delta \sum\limits_{j=L'}^{L-1} \left(\fermion_{j+1}^\dag \fermion_j^\dag + \fermion_j \fermion_{j+1}\right) -\\  \mu \sum\limits_{j=1}^{L'} \fermion_j^\dag \fermion_j - \mu' \sum\limits_{j=L'+1}^{L} \fermion_j^\dag \fermion_j,
\end{multline}
where $\Delta$ and $\mu'$ are the order parameter and chemical potential in
the topological superconductor, respectively. Here $|\mu'| < 2t$ corresponds to the topological superconductor and $|\mu'| > 2t$ to the trivial one.
The local density of states of this system is shown in Figs.~\ref{fig:connection}(c,d).
In particular, for an interface between the interacting model and a topological superconductor,
no resonance is expected at the junction, as the Majorana zero mode of the topological
superconductor and the Majorana-like mode of the interacting model
will annihilate each other (Figs.~\ref{fig:connection}(c)). In contrast, for the interface
between the interacting model and a conventional superconductor,
a zero-mode will remain at the interface between the two systems (Figs.~\ref{fig:connection}(d)).
Note that the very same phenomenology would be observed if the interacting model
is replaced by a symmetry broken topological superconductor. 

\section{Bosonized continuum limit}
\label{sec:bos}

\begin{figure}
    \centering
    \includegraphics[width=\linewidth]{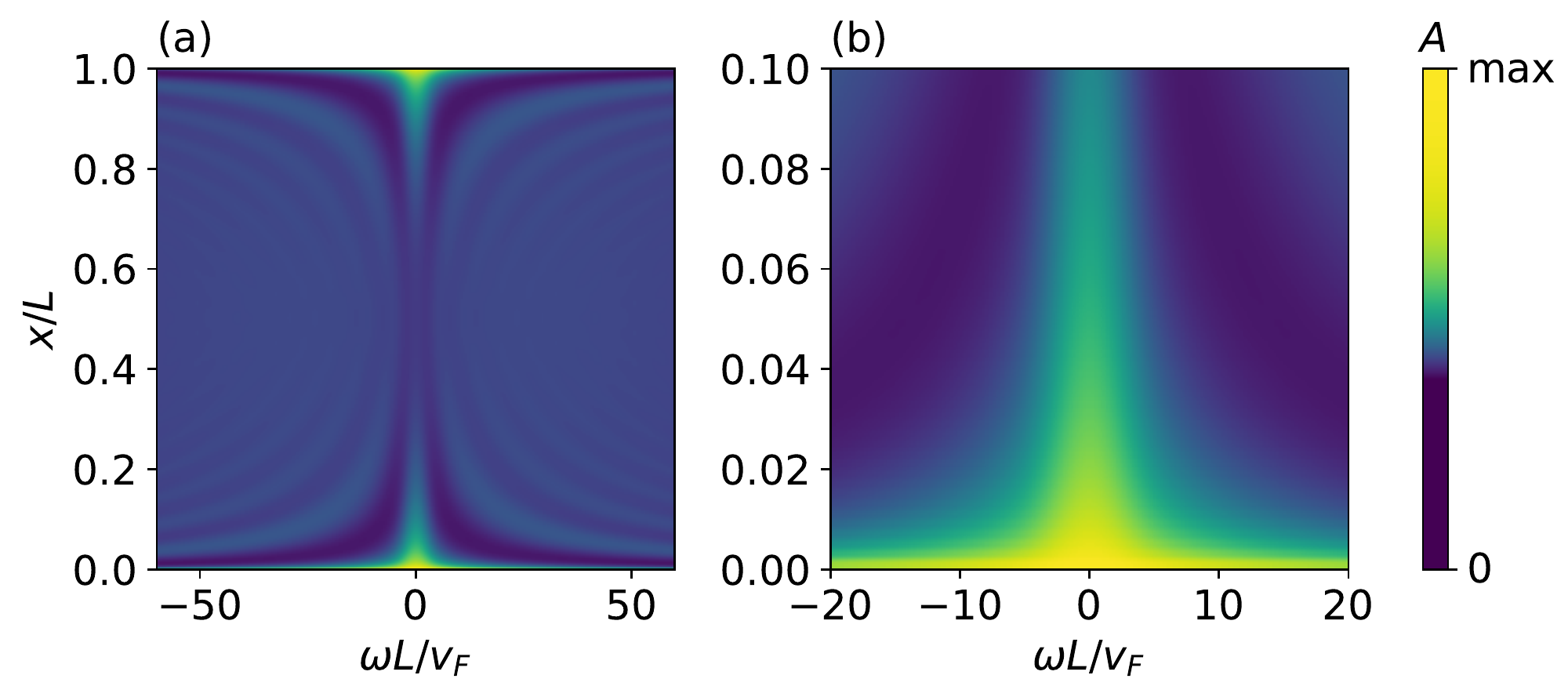} 
    \caption{Local density of states for the continuous wire of interacting fermions normalized on the local density of states of the non-interacting system. (a,b) Color map of $A(x, \omega)$,
    in the whole chain (a) and close to the edge (b). 
    Here $v_F = g$, the frequency broadening of the local density of states is equal to $\delta = v_F / (2L).$}
    \label{fig:boson}
\end{figure}

In order to show that the peaks exist for an arbitrarily weak attractive interaction we employ bosonization technique for continuous analogue of the model, described by the following Hamiltonian:
\begin{equation}
    \hat H_c = \int\limits_0^L \left\{ \hat \psi^\dag \left(\frac{\partial_x^2}{2m} - \frac{k_F^2}{2m}\right) \hat \psi + \frac{g}{8 k_F^2} : \left[\partial_x \hat \rho\right]^2:\right\}\;dx
\end{equation}
where $\hat \psi^\dag$ and $\hat \psi$ are the fermionic creation and annihilation operators, $m$ is the mass of the fermion, $k_F$ is the Fermi momentum, $g$ is the interaction parameter, $\hat \rho = \hat \psi^\dag \hat \psi$ is the density operator and $:\ldots:$ denotes the normal ordering of the operator. Here we impose zero boundary conditions $\hat \psi(0) = \hat \psi(L) = 0$. Following Ref.~\cite{fabrizio1995} we introduce an auxiliary right moving field $\hat \psi_R(x)$ defined on the segment $[-L, L]$:
\begin{equation}
    \hat \psi(x) = e^{i k_F x} \hat \psi_R(x) - e^{-ik_F x} \hat \psi_R(-x) .
\end{equation}
Zero boundary conditions for $\hat \psi(x)$ are equivalent to periodic ones for $\hat \psi_R(x)$ and, hence, the latter field can be straightforwardly bosonized. The Hamiltonian expressed in terms of $\hat \psi_R$ has the following form:
\begin{equation}
    \hat H_c \approx \int\limits_{-L}^L \left[
    -i v_F \hat \psi_R^\dag \partial_x \hat \psi_R - \frac{g}{2} : \hat \rho_R(x) \hat \rho_R(-x):
    \right]\;dx,
\end{equation} 
where $v_F = k_F / m$, $\hat \rho_R = \psi_R^\dag \psi_R$ and we have replaced the quadratic dispersion by a linear one in the vicinity of the Fermi surface. The bosonization expression for the field $\hat \psi_R(x)$ reads~\cite{von1998bosonization}
\begin{equation}
    \hat \psi_R(x) = \frac{\hat F}{\sqrt{2\pi a}} e^{i \pi \hat N x / L} e^{i \hat \phi(x)},
\end{equation}
where $\hat N$ is a number operator for extra particles with respect to the Fermi sea state and $\hat F$ is the Klein factor $[\hat N, \hat F] = -\hat F$, $\hat F^\dag \hat F = \hat F \hat F^\dag = 1$. The phase field is given by the following expression:
\begin{equation}
    \hat \phi(x) = i \sum\limits_{k>0} \sqrt{\frac{\pi}{k L}} \left(e^{-ikx} \hat b_k^\dag - e^{ikx} \hat b_k \right) e^{-ka / 2},
\end{equation}
where $k = \pi j / L$, $j$ is a positive integer, $\hat b_k$ and $\hat b_k^\dag$ are the canonical bosonic operators. We also introduce a regularization parameter $a$ and all the expressions are to be understood in the limit $a \to +0$. In terms of the bosonic operators the Hamiltonian is expressed as follows:
\begin{multline}
    \hat H_c = \sum\limits_{k>0} k \left[
    v_F \hat b_k^\dag \hat b_k - \frac{g}{4\pi} \left(\hat b_k \hat b_k + \hat b_k^\dag \hat b_k^\dag\right)
    \right] + \\ \frac{\pi}{2L} \left(v_F -  \frac{g}{2\pi}\right) \hat N^2 .
    \label{eq:bose-hamiltonian}
\end{multline}
This Hamiltonian can be easily diagonalized by the Bogoliubov transform
\begin{equation}
    \hat b_k =  \tilde b_k \cosh \varphi + \tilde b_k^\dag \sinh \varphi ,~\tanh 2 \varphi = \frac{g}{2\pi v_F}
\end{equation}
and reduced to the following form:
\begin{equation}
    \hat H_c = \sum\limits_{k>0} k \sqrt{v_F^2 - \left(\frac{g}{2\pi}\right)^2} \tilde b_k^\dag \tilde b_k + \frac{\pi}{2L}\left(v_F - \frac{g}{2\pi}\right) \hat N^2 .
    \label{eq:bose-hamiltonian-diagonal}
\end{equation}
We note that the Bogoliubov transform can be performed only if $|g| < 2\pi v_F$, otherwise the Hamiltonian ~\eqref{eq:bose-hamiltonian} is not bounded from below. The case of strong interaction corresponds to the phase separation or charge density wave which cannot be described with the bosonization formalism.

The local density of states of such a wire is given by the following expression:
\begin{equation}
    A(x, \omega) = i\int\limits_{-\infty}^{\infty} \left[G^{>}(x, x, t) - G^{<}(x, x, t)\right] e^{i \omega t}\;dt, 
\end{equation}
where
\begin{multline}
    G^{>}(x, x', t) = -i\langle \hat \psi(x, t) \hat \psi^\dag(x', 0) \rangle \approx \\
    G_R^{>}(x, x', t) + G_R^>(-x, -x', t);
    \label{eq:green-greater}
\end{multline}
\begin{equation}
    G_R^>(x, x', t) = -i\langle \hat \psi_R(x, t) \hat \psi_R^\dag(x', 0) \rangle,
\end{equation}
and 
\begin{multline}
    G^{<}(x, x', t) = i \langle \hat \psi^\dag(x', 0) \hat \psi(x, t) \rangle \approx\\
    G_R^<(x,x', t) + G_R^<(-x, -x', t),
    \label{eq:green-lesser}
\end{multline}
\begin{equation}
    G_R^<(x, x', t) = i\langle \hat \psi_R^\dag(x', 0) \hat \psi_R(x, t) \rangle.
\end{equation}
Here, in the right hand side of Eqs.~\eqref{eq:green-greater} and~\eqref{eq:green-lesser} we neglect the terms oscillating as $e^{\pm 2ik_F (x+x')}$. The Green's function of the field $\hat \psi_R$ can be evaluated analytically, see Appendix~\ref{sec:green} for details. Figure~\ref{fig:boson} shows the local density of states of the wire as a function of coordinate and frequency. One can clearly see the peaks at the ends of the wire at zero frequency. 

The form~\eqref{eq:bose-hamiltonian-diagonal} of the Hamiltonian allows to evaluate the gap in the local density of states. This gap is equal to $\pi / (2L) [v_F - g / (2\pi)]$ since it is the minimal energy needed to add or remove a single particle to or from the wire without excitation of the bosonic modes. This expression is in the good qualitative agreement with the results shown in the Fig.~\ref{fig:scaling}: the peak splitting scales as $C/L$ and the factor $C$ linearly decays with the increase of interaction up to the transition to the phase separation.

\section{Summary}
\label{sec:con}

To summarize, we have demonstrated the emergence of zero-bias modes at the edges of a 1D chain of attractively
interacting fermions. We have demonstrated that these modes are adiabatically connected
to conventional Majorana zero modes of a topological
superconductor, yet with the striking difference that they emerge in a situation without gauge symmetry breaking.
These many-body zero modes are found to exist at arbitrarily weak attractive interaction
and are robust to perturbations which break the integrability of the model, 
including long-range hopping, interactions, and disorder. 
In particular, we have demonstrated that these zero-mode resonances can be rationalized both with
lattice quantum many-body formalism based on tensor networks, and with a continuum low energy model
based on bosonization. Our results put forward a new type of protected zero modes in a purely many-body limit,
with no single-particle analog, providing a stepping stone towards the exploration
of topological modes in generic quantum disordered many-body systems.

\acknowledgments
We acknowledge the financial support from our Academy of Finland projects (Grant Nos. 331342 and 336243) and its Centre of Excellence in Quantum Technology (QTF) (Grant Nos. 312298 and No. 312300), and from the European Research Council under Grant No. 681311 (QUESS). 
The research was also partially supported by the Foundation for Polish Science through the IRA Programme co-financed by EU within SG OP. 
We thank Sergei Sharov and Pavel Nosov for useful discussions and the Aalto Science-IT project for computational resources.

\bibliography{bibliography}

\appendix

\section{Critical point}
\label{sec:critical}

In the general interacting case the Bethe ansatz expression for the ground state of the system is complicated except for the critical point $\mu = 0$, $V = 2t$. In this case the ground state has the form~\cite{casiano2019quantum}
\begin{equation}
    |\Psi_0\rangle = \frac{1}{2^{L/2}} \prod_{j=1}^L \left(1 + \fermion_j^\dag\right)|0\rangle
\end{equation}
corresponding to energy $E_0 = t (1-L)/2$. Interestingly,
this is actually same ground state as in Kitaev model, and thus
it would be possible to
construct Majorana operators describing localized zero-energy excitations~\cite{Greiter2014}. It must be noted
that in the present model there will be also additional low-energy excitations as we are considering a
truly interacting model instead of mean-field Hamiltonian.

In order to prove this result we define $|\Psi_{j-1}\rangle$ be defined as
\begin{equation}
    |\Psi_{j-1} \rangle = \frac{1}{2^{L/2}}\prod\limits_{k=j}^L\left(1 + \fermion_k^\dag\right)|0\rangle .
\end{equation}
At the same time we define $\hat H_{j-1}$ as:
\begin{multline}
    \hat H_{j-1} = -t \sum\limits_{k=j}^{L-1} \left(\fermion_{k+1}^\dag \fermion_k +
    \fermion_k^\dag \fermion_{k+1}\right) -\\  2t \sum\limits_{k=j}^{L-1} \left(\hat
    \fermion_{k+1}^\dag \fermion_{k+1} - \frac{1}{2}\right)\left(\fermion_{k}^\dag \hat
    \fermion_k - \frac{1}{2}\right).
\end{multline}
It's obvious that $\hat H_0 = \hat H_I$ at $V = 2t$ and $\mu = 0$. We
are going to prove that $|\Psi_{j-1}\rangle$ is the eigenstate of the Hamiltonian
$\hat H_{j-1}$ with the energy $t(j  - L)/2$. For $j=L$ this fact is obvious because
$\hat H_{L-1} = 0$. Assume that for some $j-1$ this statement is true. Then let us
prove it for $j - 2$:
\begin{widetext}
\begin{multline}
    \hat H_{j-2} |\Psi_{j-2}\rangle = \left[-t \left(\fermion_{j}^\dag \hat
    \fermion_{j-1} + \fermion_{j-1}^\dag \fermion_j\right) - 2t\left(\fermion_{j}^\dag \hat
    \fermion_{j} - \frac{1}{2}\right)\left(\fermion_{j-1}^\dag \fermion_{j-1} -
    \frac{1}{2}\right) + \hat H_{j}\right]\left(1 + \fermion_{j-1}^\dag \right)
    |\Psi_{j-1}\rangle = \\
    \frac{t(j  -L)}{2} |\Psi_{j-2}\rangle -t \left[
        \fermion_{j-1}^\dag |\Psi_{j}\rangle - \left(\fermion_{j}^\dag \hat
        \fermion_{j} - \frac{1}{2}\right) |\Psi_{j-1}\rangle +
        \fermion_{j}^\dag \fermion_{j-1} \fermion_{j-1}^\dag |\Psi_{j-1}\rangle +
        \right. \\ \left.
        2\left(\fermion_{j}^\dag \hat
        \fermion_{j} - \frac{1}{2}\right)\left(\fermion_{j-1}^\dag \fermion_{j-1} -
        \frac{1}{2}\right) \fermion_{j-1}^\dag |\Psi_{j-1}\rangle
        \right] =  \\
    \frac{t(j  -L)}{2} |\Psi_{j-2}\rangle - t \left[
        \fermion_{j-1}^\dag |\Psi_{j} \rangle - \fermion_j^\dag
        |\Psi_{j}\rangle + \frac{1}{2} |\Psi_{j-1}\rangle + \fermion_j^\dag
        |\Psi_{j}\rangle + \left(\fermion_j^\dag \fermion_j -
        \frac{1}{2}\right)\fermion_{j-1}^\dag |\Psi_{j-1}\rangle
        \right] = \\
    \frac{t(j -L)}{2} | \Psi_{j-2}\rangle -t \left[
        \fermion_{j-1}^\dag |\Psi_{j}\rangle + \frac{1}{2} |\Psi_{j-1}\rangle +
        \fermion_{j-1}^\dag \fermion_{j}^\dag |\Psi_{j}\rangle - \frac{1}{2}
        \fermion_{j-1}^\dag |\Psi_{j - 1}\rangle
        \right] = \\
    \frac{t(j -L)}{2} |\Psi_{j-2}\rangle -t \left[
        \frac{1}{2} | \Psi_{j - 1}\rangle + \frac{1}{2} \fermion_{j-1}^\dag |\Psi_{j-1}\rangle
        \right] = \frac{t(j-1 - L)}{2} |\Psi_{j-2}\rangle .
\end{multline}
\end{widetext}
In order to prove that the found eigenstate is the ground state, i.e. has the minimal
possible energy, we notice that the Hamiltonian can be written as:
\begin{multline}
    \hat H_I = -t \sum\limits_{j=1}^{L-1} \left[
        \fermion_{j+1}^\dag \fermion_j +
        \fermion_j^\dag \fermion_{j+1} + \phantom{\frac 12}\right. \\ \left.
        2 \left(\fermion_{j+1}^\dag \fermion_{j+1} - \frac{1}{2}\right)
        \left(\fermion_j^\dag \fermion_j - \frac{1}{2}\right)
        \right]
\end{multline}
Each term of the sum has the lowest eigenvalue equal to $-t / 2$ and there are
$L-1$ terms total, then the energy of the ground state cannot be lower than
$t(1 - L) / 2$.

Since the state $|\Psi_0\rangle$ is not an eigenstate of the number of particles operator, one can project it onto the subspaces with the fixed number of particles yielding $L+1$-times degenerate ground states corresponding to the different number of particles $N$ from $0$ to $L$:
\begin{equation}
    |\Psi_{0N}\rangle = \frac{2^{L/2}}{\sqrt{L\choose N}} \hat P_N |\Psi_0\rangle,
\end{equation}
where $\hat P_N$ is an orthogonal projector onto the subspace with $N$ particles.

\section{Green's functions of the continuous wire}
\label{sec:green}

The greater and lesser Green's functions of the field $\hat \psi_R$ are equal to
\begin{multline}
    G_R^>(x, x', t) = \\ -\frac{i}{2\pi a} \left\langle
    \hat F(t) e^{i \pi \hat N (x - x') / L} \hat F^\dag e^{i \hat \phi(x,t)} e^{-i\hat \phi(x')}
    \right\rangle;
\end{multline}
\begin{multline}
    G_R^<(x, x', t) = \\ \frac{i}{2\pi a} \left\langle
    e^{-i \pi \hat N x' / L} \hat F^\dag  \hat F(t) e^{i \pi \hat N x / L} e^{-i \hat \phi(x')} e^{i\hat \phi(x, t)}
    \right\rangle,
\end{multline}
where $\hat F(t) = e^{i \hat H_c t} \hat F e^{-i \hat H_c t}$ and $\hat \phi(x, t)= e^{i \hat H_c t} \hat \phi(x) e^{-i\hat H_c t}$. After a long but straightforward calculations one obtains:
\begin{widetext}
\begin{multline}
    G_R^>(x, x', t) = 
    -\frac{i c}{2L} \exp\left[- \frac{i\pi t}{2L}\left(v_F - \frac{g}{2\pi}\right) + \frac{ i\pi (x - x')}{L}\right] \times \\ \frac{
        \left[1 - e^{-\frac{\pi}{L} (a + i\tilde v_F t + i x + i x')}\right]^{\frac s2}
        \left[1 - e^{-\frac{\pi}{L} (a + i\tilde v_F t - i x - i x')}\right]^{\frac s2}
        \left[1 - e^{-\frac{\pi}{L} (a + i\tilde v_F t + i x - i x')}\right]^{\frac{1 -c}{2}}
    }{
        \left|1 - e^{-\frac{\pi}{L} (a - 2 i x)}\right|^{\frac s2}
        \left|1 - e^{-\frac{\pi}{L} (a - 2 i x')}\right|^{\frac s2}
        \left[1 - e^{-\frac{\pi}{L} (a + i \tilde v_F t - i x + i x')}\right]^{\frac{1 + c}{2}}
    };
\end{multline}
\begin{multline}
    G_R^<(x, x', t) = \frac{i c}{2L} \exp\left[ \frac{i\pi t}{2L}\left(v_F - \frac{g}{2\pi}\right)\right] \times \\ \frac{
        \left[1 - e^{-\frac{\pi}{L} (a - i\tilde v_F t - i x - i x')}\right]^{\frac s2}
        \left[1 - e^{-\frac{\pi}{L} (a - i\tilde v_F t + i x + i x')}\right]^{\frac s2}
        \left[1 - e^{-\frac{\pi}{L} (a - i\tilde v_F t - i x + i x')}\right]^{\frac{1 -c}{2}}
    }{
        \left|1 - e^{-\frac{\pi}{L} (a + 2 i x)}\right|^{\frac s2}
        \left|1 - e^{-\frac{\pi}{L} (a + 2 i x')}\right|^{\frac s2}
        \left[1 - e^{-\frac{\pi}{L} (a - i \tilde v_F t + i x - i x')}\right]^{\frac{1 + c}{2}}
    },
\end{multline}
\end{widetext}
where $c = \cosh 2 \varphi$, $s = \sinh 2 \varphi$ and $\tilde v_F = \sqrt{v_F^2 - g^2 / (2\pi)^2}$. Finally, the expression for the local density of states reads as:
\begin{widetext}
\begin{equation}
    A(x, \omega) = 2 \int\limits_{-\infty}^{+\infty} e^{i\omega t} \mathop{\mathrm{Re}} \left\{e^{-\frac{i \pi t}{2 L}\left(v_F - \frac{g}{2\pi}\right)} \frac{\left(1 - e^{-\frac{\pi a}{L}}\right)^c
    \left[1 - e^{-\frac{\pi}{L} (a - i \tilde v_F t - 2 i x)}\right]^{\frac s2}
    \left[1 - e^{-\frac{\pi}{L} (a - i \tilde v_F t + 2 i x)}\right]^{\frac s2}
    }{\pi a \left |1 - e^{-\frac{\pi}{L} (a + 2 i x)}\right|^s
    \left[1 - e^{-\frac{\pi}{L} (a - i v_F t)}\right]^c
    }\right\}\;dt.
\end{equation}
\end{widetext}

\section{Persistent current in a ring}
\label{sec:persistentcurrent}

\begin{figure}
    \centering
    \includegraphics[width=\linewidth]{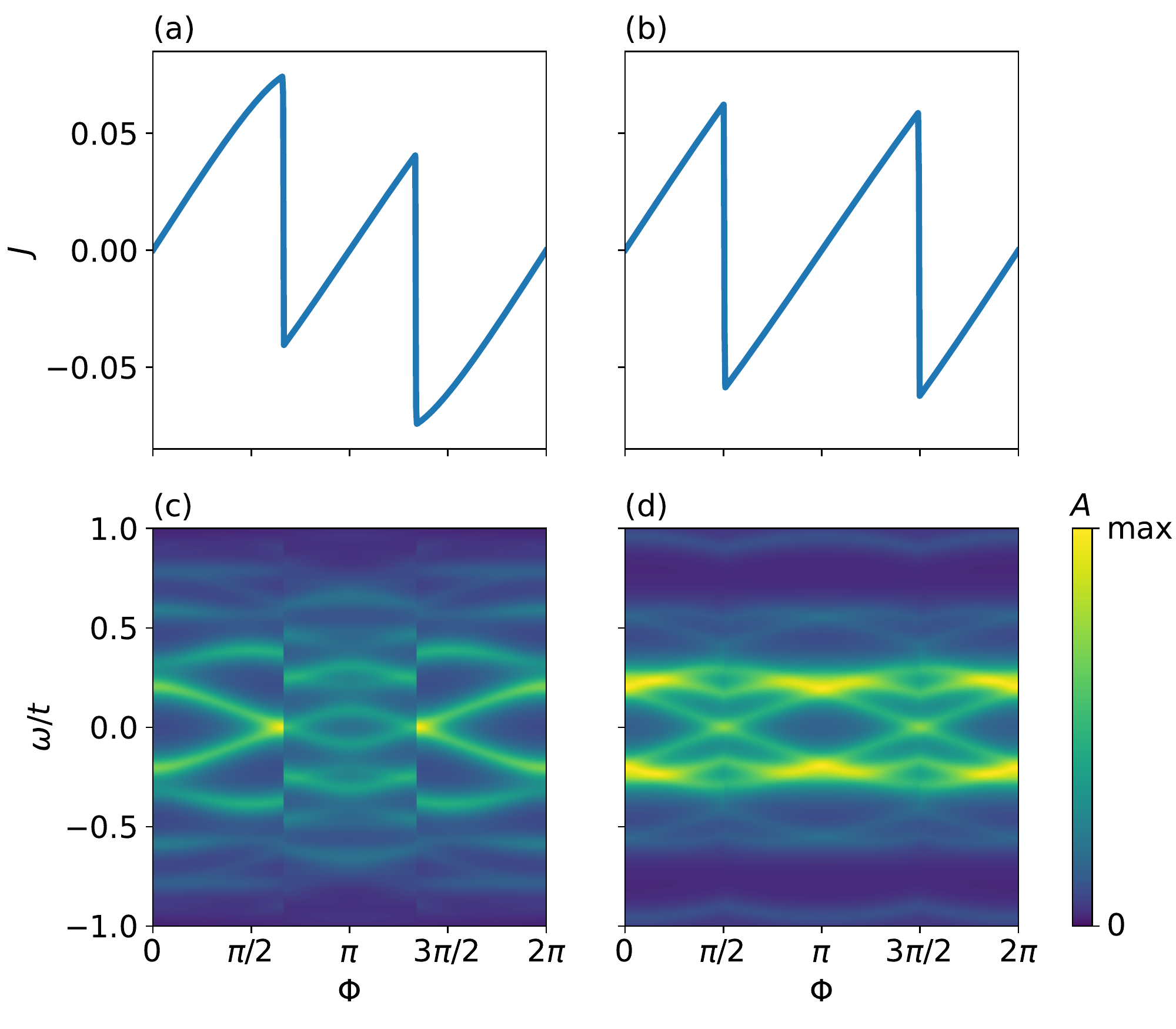}
    \caption{(a), (b) The persistent current in the ring as a function of flux.  (c), (d) The spectral function at the site adjacent to the weak link as a function of frequency and flux. The parameters are $t_w = 0.2 t$, $L = 10$, $\mu = 0$, (a), (c) $V = 1.5 t$, (b), (d) $V = 2 t$. The data is obtained using exact diagonalization of the Hamiltonian~\eqref{eq:jj}. The frequency broadening of the local density of states is equal to $\delta = 0.05 t$.}
    \label{fig:jj}
\end{figure}

We close up the chain into a ring with a weak link and pierce it with a magnetic flux. The Hamiltonian of such a system reads as:
\begin{equation}
    \hat H_I' = \hat H_I - t_w \left(\fermion_{L}^\dag \fermion_1 e^{i \Phi} + \fermion_1^\dag \fermion_L e^{-i\Phi}\right),
    \label{eq:jj}
\end{equation}
where $t_w$ is hopping parameter through the weak link between the first and the last sites of the chain, and $\Phi$ determines the flux in the units of $\Phi_0 / (2\pi)$, where $\Phi_0$ is the normal (not superconducting) flux quantum. The current operator is defined as
\begin{equation}
    \hat J = \frac{\partial \hat H'_I}{\partial \Phi} = -i t_w \left(\fermion_{L}^\dag \fermion_1 e^{i \Phi} - \fermion_1^\dag \fermion_L e^{-i\Phi} \right).
\end{equation}
The current dependence on the flux is shown in Fig.~\ref{fig:jj}(a), (b) for different values of interaction strength. The discontinuities in the current-flux relation correspond to the number of particles switches in the ring. Figures~\ref{fig:jj}(c) and (d) show the local density of states at the site adjacent to the weak link as a function of flux and frequency. The peak splitting oscillates as a function of flux, the number of particles switches take place exactly at the peak's intersections. The behavior of the current and spectral function at the critical point $V = 2t$ mostly resembles the behavior of the mean field Kitaev model where these quantities are $\pi$-periodic with the flux if one allows parity switches, and $2\pi$ periodic if the parity is kept fixed. Away from the critical point we do not observe exact $\pi$-periodicity of the current in the presence of parity switches  which may be a consequence of a finite overlap of the peaks through the bulk of the chain. \

\end{document}